\documentclass[12pt,a4paper]{article}
\usepackage{epsfig}
\begin{document}

\begin{titlepage}

\begin{center}
\Huge

The detective Modulation Transfer Function applied to scanned digital
radiographic detectors.

\vspace{1.5cm}

\Large 
D.J. Miller$^{*}$ and A. Papanestis

\large
University College London, UK

\small

\vspace{1cm}
\begin{flushleft}
$^{*}$Corresponding author\\
Dept. of Physics \& Astronomy\\
University College London\\
Gower St.\\
London WC1E 6BT\\
UK

\vspace{3mm}
email: djm@hep.ucl.ac.uk\\
tel: +44 171 3807152\\
fax: +44 171 3807145 
\end{flushleft}
\end{center}

\renewcommand{\baselinestretch}{1.5}
\normalsize

\begin{center}
\vspace{2cm}

{\bf Abstract}
\end{center}

An extension of the traditional modulation transfer function (MTF), the
detective MTF, is proposed for the evaluation of the resolution of 
X-ray imaging detectors with non-uniform efficiency on the scale of individual
detector elements. The dMTF is shown 
to give a better representation than does the MTF of the improved 
resolution reported when the readout of the detectors is 
enhanced.

\end{titlepage}

\renewcommand{\baselinestretch}{1.5}
\large
\normalsize

\section{Introduction}

The resolution of an imaging detector is frequently expressed in terms
of a modulation transfer function (MTF); the Fourier transform of the
line spread function (LSF) in one dimension, or of the point spread
function (PSF) in two dimensions \cite{med}.  But for certain kinds of
detector \cite{Lev2} it has been noticed that changes in the operating mode of
the readout electronics can bring clear improvements to the quality of
the image without any significant change in the MTF, as normally
defined.  A simple modification to the LSF is suggested, giving a
``detective line spread function'' (dLSF), the Fourier transform of
which is the detective modulation transfer function (dMTF).  For the
detectors studied the  resulting measure of the resolution is shown to
be significantly improved when the readout mode is changed, matching
the results of other methods of assessing the resolution of the
system.

\section{Detectors and Readout}

The Siberian Digital Radiographic Device (SDRD) \cite{Lev2,Lev1,Lev3}
detects X-rays in a Xenon-filled multiwire proportional chamber
working at high pressure.  It is a linear array of independent
sub-detectors, with 320 separate sense wires, 5~cm long and
approximately 1~mm apart.  To eliminate parallax the wires are aligned
along the local direction of the X-ray beam, giving an increase of
spacing from 1.20 to 1.25~mm  over the thickness of the chamber.  Each
wire is connected to its own pre-amplifier, which is followed by
thresholding and logic circuits that feed into parallel scalers. Every
15 or 30~milliseconds (depending on the version of the device) the
contents of the whole set of scalers are transferred to memory to give
a single scan line.  The detector, the X-ray source and the
collimators which limit acceptance are all mounted on a rigid gantry
which is scanned transversely  at a rate of about 10~cm/sec, building
up a 2-dimensional picture which is stored in RAM. 

In the original low resolution standard (ST) version of the device
there is one scaler associated with each sense-wire.  Because of the
finite size of the charge cluster produced by a single X-ray, a
significant fraction of all hits gives pulses above threshold on two
adjacent channels.  A  coincidence circuit is provided between each
pair of channels to inhibit counts from these double hits.  Only
channels with single hits are counted for the image. 

In the high resolution (HR) version a second set of scalers is
provided, one between each pair of wires.  Single hits are counted in
the wire scalers, as before, but double hits are counted in the extra
coincidence scalers, giving a much higher efficiency.  The HR system
has been shown to resolve significantly finer detail than the ST
system, both in laboratory tests \cite{Arni} and in clinical use \cite{Lev4}. 

The present work is motivated by the development of a new imaging
system, optimised for mammographic examinations, which will use a
similar scanning geometry and electronics to the SDRD, but with a
different kind of  detector, a microstrip gaseous chamber (MSGC) with
a strip pitch of approximately 200 micrometres.  Although the
intrinsic resolution of the MSGCs is expected to be a factor of $\sim$1/5
of that in the SDRD, the response function of each strip has a very
similar shape. 

Laboratory tests have been made on test MSGCs with parallel strips,
both with Xenon at high pressure and with Argon at atmospheric
pressure.  On the basis of these tests a computer simulation has been
developed which can be used to predict the performance of a high
pressure MSGC with  the strips tapered towards the X-ray source in the
same way as in the SDRD (similar tapered-strip MSGCs are also being
developed for the CMS detector at the CERN Large Hadron Collider \cite{CMS}).

\section{Resolutions and Efficiencies}

In Figure~1 curve a) shows the simulated response of one MSGC
strip when a narrow X-ray source  is scanned across it.  This
represents the line spread function of the linear detector.   If the
same kind of coincidence electronics is used as in the SDRD, then the
LSF is sharpened by the suppression of double hits, as shown by curve b).
And if a second set of scalers is provided, as in the HR
mode of the SDRD, then an additional LSF can be obtained, representing
the response of the coincidence channel as the narrow source is
scanned across the detector (curve c).   The RMS spreads of the  LSFs are
shown in the same figure. 

Just as with the SDRD, it is possible to consider operating the MSGC
device in either the ST mode, with only single hits on strips counted
and used in the image, or in HR mode, with an extra set of scalers for
the double hits.  Figure~2 shows the efficiency of the detector in
registering hits from X-ray photons that have converted in its active
area $\eta(x)$.  Clearly the HR mode makes much better use of the X-rays
which reach the detector, hence minimising the dose to the patient.
(It appears from Figure~2 that the HR mode can sometimes exceed very
slightly 100\% efficiency. This is due to the
few events in which 3 channels are hit giving two
apparent hits from a single event. There is no significant effect on
the argument given here.) 

The LSFs shown in Figure~1 for the single strip hits and for the
coincidences are very similar in width.  But an apparent paradox
arises if the normal definition of MTF is used to estimate the
effective resolution of the chamber in the two modes. In the ST mode
one would use the LSF of curve b) on Figure~1.  In the HR mode
it would be reasonable to take the average of curves b) and c)
as the response of the detector (Figure~1 curve d), which 
means that the predicted MTFs in
the two modes would be  almost identical.  As mentioned in section~2 above,
if this is taken to mean identical image quality it contradicts both
laboratory measurements and clinical experience with the SDRD.

\section{Definition of the dLSF and the dMTF}

The detective line spread function is defined as:
\begin{equation}
dLSF(x)=LSF(x)/\eta (x)	
\label{eq:dLSF}
\end{equation}
For the ST system, with rejection of coincidences, the LSF as a
function of x from Figure~1, curve b), is divided by the efficiency
$\eta(x)$ from Figure~2, giving a fatter curve for the dLSF
shown as circles in Figure~3 a), (for comparison, the
stars represent the unmodified LSF from Figure~1).  The corresponding
dMTF is obtained by taking the Fourier transform of the dLSF (Figure~3 b).
The 10\% level on the dMTF gives a resolution of 4.5 line
pairs per mm, whereas the unmodified MTF would suggest a resolution of
6.5 line pairs per mm.  

In the HR mode the combined relative efficiency of the two kinds of
channel remains close to $\eta(x)=1$ for all x, so there is no visible
difference between the LSF and  the dLSF, shown in Figure~3 c), or
between the MTF and the dMTF, shown in Figure~3 d).  The resolution
at the 10\% level of  either the MTF or the dMTF is 6.5 line pairs per
millimetre. 

The detailed shapes of the dLSF and dMTF curves in Figure 3 show
that the response of the ST mode resembles the response of a detector
with a 200~$\mu$m square aperture, while that of the HR mode 
resembles a Gaussian response with a FWHM of 138~$\mu$m.

\section{Conclusions}

The detective line spread function, defined in equation~\ref{eq:dLSF},
can be Fourier transformed to give the detective modulation transfer
function (dMTF) which is a more realistic measure of the true
resolution for a detector whose efficiency varies significantly with
position within the scale of variations of the LSF.  Using the dMTF it
is possible to give an objective prediction of the improved
performance which can be expected from a Microstrip Gaseous Detector
if it is operated in the HR mode, with separate scalers for
single-strip hits and for coincidences between adjacent wires.  With a
strip pitch of approximately 200~$\mu$m the resolution in HR mode is
6.5~line pairs per millimetre  whereas, in standard (ST) mode, the
resolution would be 4.5 line pairs per millimetre due to significant
reductions in efficiency midway between the strips.  This result
explains the clear improvement in image quality which has been
reported for the Siberian Digital Radiographic Device \cite{Arni} when 
it was upgraded from the ST to the HR mode. 

\section{Acknowledgements}

This work would not have been possible without the collaboration and
hospitality of L.I.~Shekhtman and A.G.~Khabakhpashev of the Budker
Institute of Nuclear Physics in Novosibirsk, Russia, or the advice of
R.D.~Speller of the UCL Medical Physics Department.  Support was
received from the  UK Particle Physics and Astronomy Research Council
and from the European Union Radiation Protection programme.

\vspace{1cm}
{\bf Figure~1.} The simulated LSFs of an MSGC operating at different modes:
a) single strip without special electronics; b) single hits; c) 
coincidences; d) average of b) and c).
\vspace{2mm}

{\bf Figure~2.} Comparison of the efficiency of the detector in
registering hits from X-ray photons that have converted in its active
area for the ST and HR modes.
\vspace{2mm}

{\bf Figure~3.} The LSF compared with the dLSF, and the MTF compared
with the dMTF for the ST and HR modes.

\topmargin-1cm
\textheight24cm

\pagebreak

\begin{center}
\epsfig{file=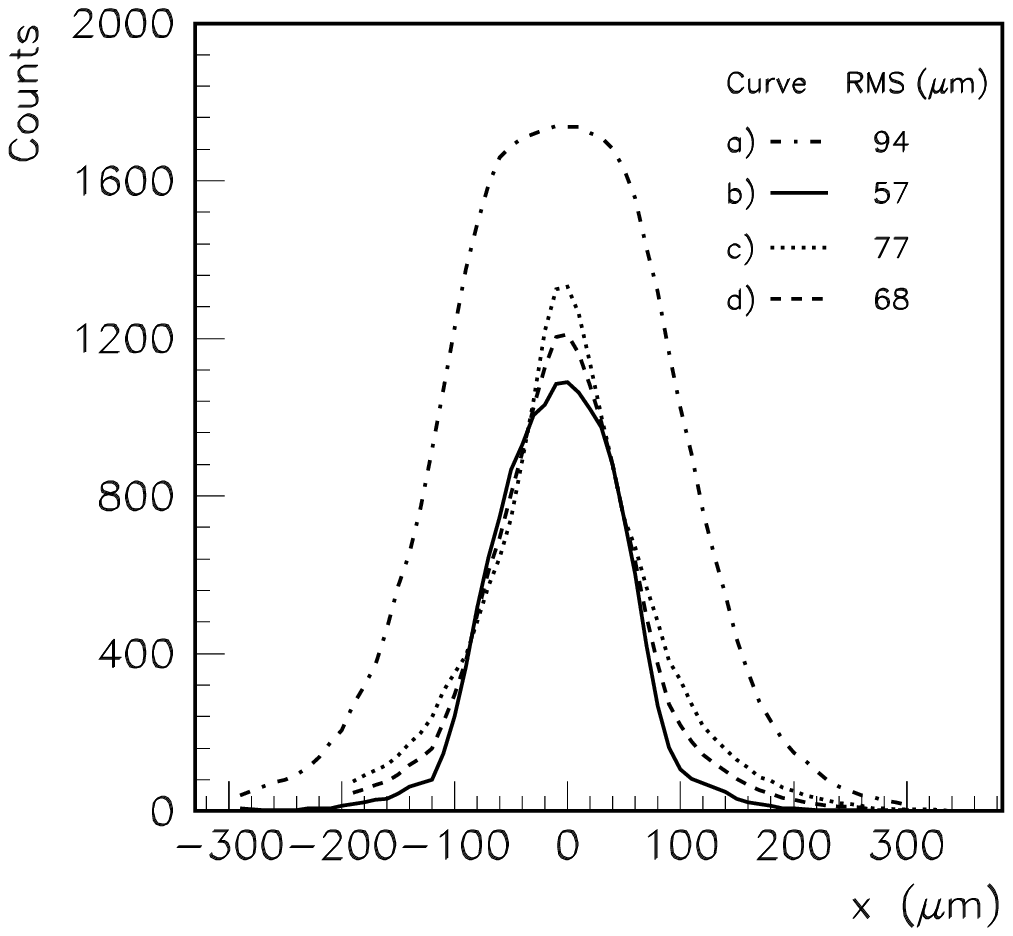}
\end{center}
{\bf Figure~1.} The simulated LSFs of an MSGC operating at different modes:
a) single strip without special electronics; b) single hits; c) 
coincidences; d) average of b) and c).

\begin{center}
\epsfig{file=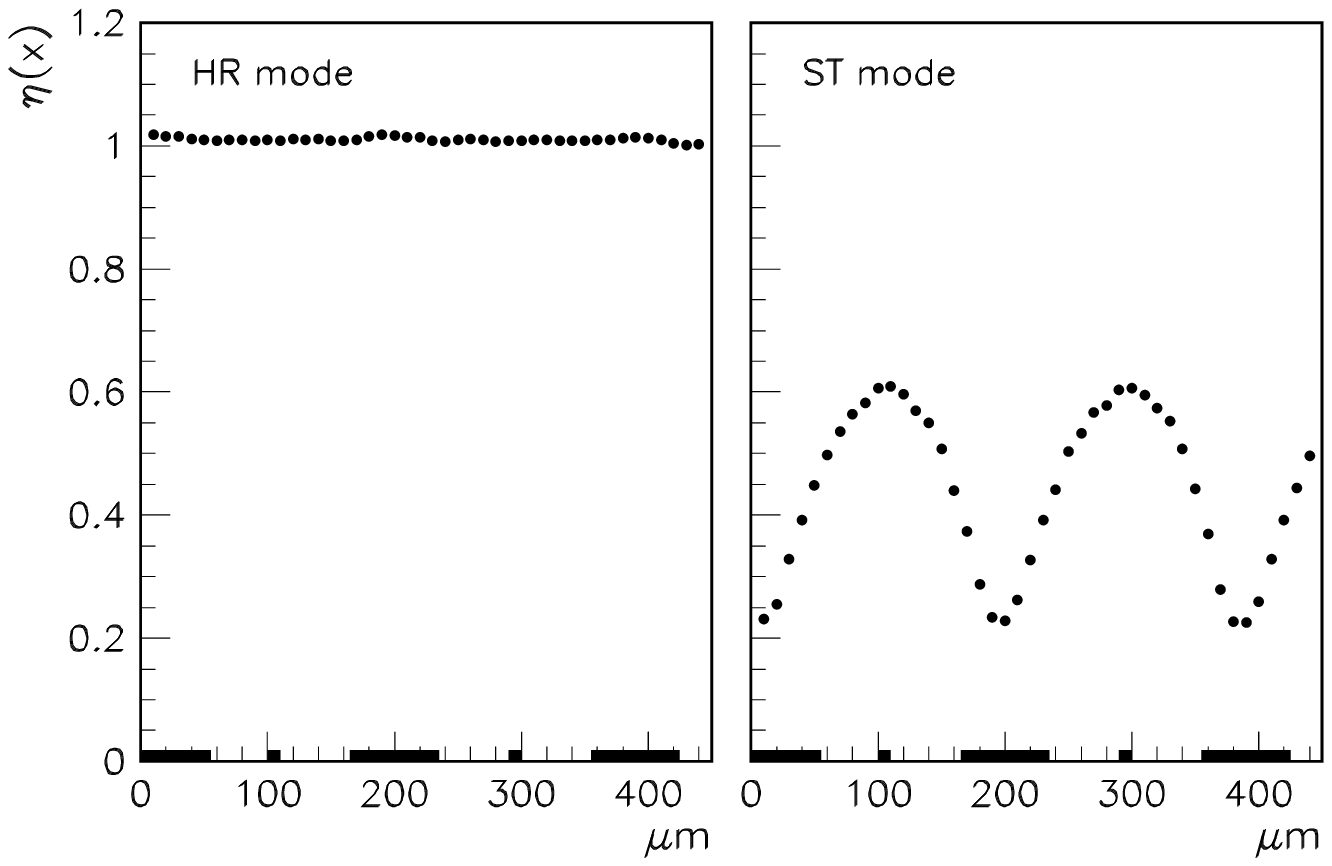,height=8cm,width=14cm}
\end{center}
{\bf Figure~2.} Comparison of the efficiency of the detector in
registering hits from X-ray photons that have converted in its active
area for the ST and HR modes.

\pagebreak

\begin{center}
\hspace*{-2cm}
\epsfig{file=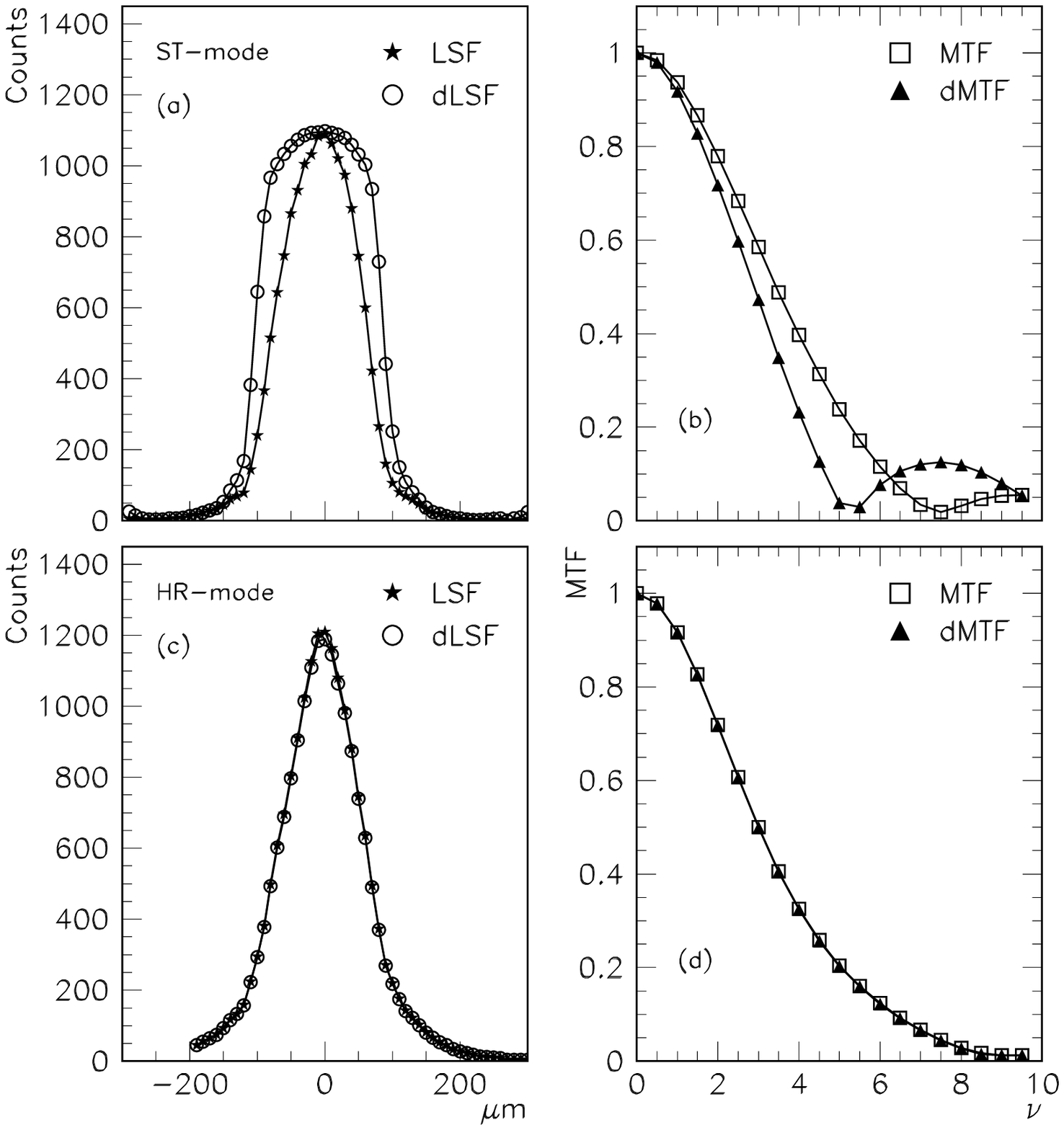}
\end{center}
{\bf Figure~3.} The LSF compared with the dLSF, and the MTF compared
with the dMTF for the ST and HR modes.

\end{document}